%% file: arxiv.tex
\documentclass[aps,prl,reprint,amssymb,showpacs,superscriptaddress,notitlepage,twocolumns]{revtex4-2}
\usepackage{bm}
\usepackage{graphicx,hyperref}
\usepackage{dcolumn}
\usepackage{braket}
\usepackage{natbib}
\usepackage{amsmath}
\usepackage{color}
\usepackage{siunitx}
\usepackage{mathtools,amssymb}
\usepackage{soul}

\begin{document}

\title{Chiral Flat-Band Optical Cavity with Atomically Thin Mirrors}

\input{authors}

\begin{abstract}
A fundamental requirement for photonic technologies is the ability to control the confinement and propagation of light. Widely utilized platforms include two-dimensional (2D) optical microcavities in which electromagnetic waves are confined between either metallic or distributed Bragg reflectors. Recently, transition metal dichalcogenides hosting tightly bound excitons with high optical quality have emerged as promising atomically thin mirrors. In this work, we propose and experimentally demonstrate a sub-wavelength 2D nano-cavity using two atomically thin mirrors with degenerate resonances. Angle-resolved measurements show a flat band, which sets this system apart from conventional photonic cavities. Remarkably, we demonstrate how the excitonic nature of the mirrors enables the formation of chiral and tunable optical modes upon the application of an external magnetic field. Moreover, we show the electrical tunability of the confined mode. Our work demonstrates a mechanism for confining light with high-quality excitonic materials, opening perspectives for spin-photon interfaces, and chiral cavity electrodynamics.
\end{abstract}

\maketitle

The ability to confine light to small volumes is central for engineering light-matter interaction in photonic and optoelectronic technologies \cite{o2009photonic,Carusotto2013,chang2014quantum,bloch2022strongly}. Planar microcavities are a key platform for confining the spatial extent of electromagnetic waves and manipulating the photonic density of states \cite{Vahala2003,Megahd2022}, which has enabled many applications such as filtering \cite{Monifi2013}, lasing \cite{Michalzik2013}, optical detection \cite{Furchi2012}, and all-optical switching \cite{Ma2006,Nakamura2004}. In these cavities, standing optical modes form between two mirrors, which are traditionally metallic or dielectric \cite{Kavokin2008}.

In addition to compactness and efficiency, another highly desirable feature for optical devices is chirality \cite{lodahl2017chiral}, a characteristic that emerges due to symmetry breaking. Recent research has focused on chiral coupling between light and emitters for classical and quantum optical applications, such as non-reciprocal optical routers and spin-photon interfaces. Examples include engineering polarization-selective spin-photon interfaces in photonic waveguides \cite{sollner2015deterministic, barik2018topological, mehrabad2023} and ring resonators \cite{barik2020chiral, jalali2020semiconductor, mehrabad2020chiral, mehrabad2023}, polaritonic chiral microcavities through magnetic \cite{Lyons2022, Suarez-Forero2023} or optical \cite{carlon2019optically} manipulation of an active medium hosted in the cavity, as well as realizing topological photonic states by time-reversal symmetry breaking \cite{klembt2018exciton}. The search for devices with non-reciprocal circular dichroism for different applications extends to other frequency domains such as infrared \cite{du2018monolithic} and terahertz \cite{mu2019tunable}.

Recently, transition metal dichalcogenides (TMDs) have emerged as a new material platform for exploring photon confinement and chiral light-matter coupling. In particular, strong and narrowband reflection has been demonstrated in monolayer MoSe$_2$ thanks to their high optical quality, i.e., a large ratio of radiative ($\Gamma_{\text{r}}$) to non-radiative ($\Gamma_{\text{nr}}$) decay rates \cite{Scuri2018,Back2018}. The integration of TMDs with photonic structures has also enabled several chiral phenomena, including spin-polarized excitons \cite{shreiner2022electrically}, hybrid exciton-polaritons \cite{liu2020generation,li2021experimental} and phonon-polaritons \cite{guddala2021topological}, by leveraging their valley-dependent optical selection rules. In these demonstrations, however, the TMDs are utilized as the active optical component instead of constituting the photonic structures \cite{Srivastava2015,Xu2014}.

\begin{figure*}
    \centering
\includegraphics[width=2\columnwidth]{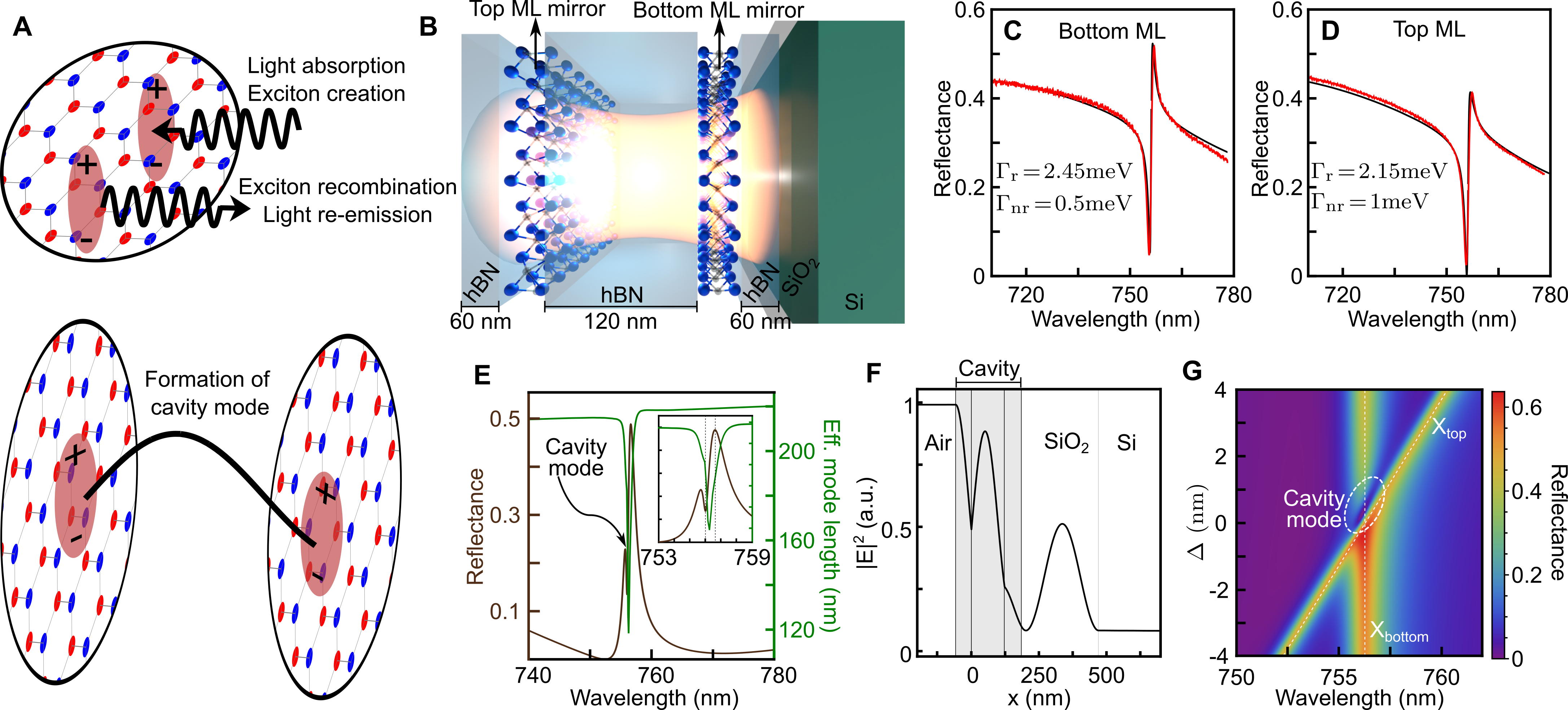}
    \caption{Design and fabrication of a cavity based on atomically thin mirrors. a) Mechanism for the realization of a nano-cavity based on atomically thin mirrors. Upper panel: Due to the high optical quality of the exciton in the material, the monolayer (ML) effectively acts as a mirror at the resonant wavelength. Lower panel: stacking two mirrors separated by dielectric material can lead to the formation of optical modes in the structure. b) Schematic representation of the TMD nano-cavity device: two atomically thin MoSe$_2$ mirrors embedded in hBN confine the electromagnetic mode. c-d) Individual reflectance spectra of the component monolayers before stacking the final device (red). The black lines show TMM fittings calculated by using a Lorentz oscillator model with the decay rates indicated in each panel. e) Simulation of the device reflectance calculated via TMM simulations. The inset shows a zoom-in of the reflectance and the corresponding Mode Effective Length calculated via FDTD, in a reduced range of wavelengths. f) Electric field intensity distribution from FDTD simulation. The minimum in both the reflectance spectrum and effective mode length (e), and enhancement of the electric field intensity profile at resonance (f) indicate the formation of a standing optical mode. g) TMM simulation of the device's spectrum upon variable exciton energy of the top monolayer $\rm{X}_{\rm top}$.  $\rm{X}_{\rm bottom}$ denotes the excitonic resonance of the bottom monolayer and $\Delta\!=\!\rm{X}_{\rm top}-\rm{X}_{\rm bottom}$. The formation of a cavity mode manifests as a minimum in the reflectance for the range of parameters indicated by the dashed white ellipse, and its energy can be manipulated by detuning the resonances of the component MoSe$_2$ mirrors.}
    \label{sketch}
\end{figure*}

In this work, we propose and experimentally demonstrate a method for realizing nanometer-thick planar optical cavities with intrinsic chiral characteristics using two atomically thin TMD mirrors as the fundamental photonic components. In contrast to conventional Fabry–Pérot interferometric cavities, the electromagnetic mode in our system arises via the efficient optical excitation and recombination of excitons in the two TMD mirrors (Fig.~\ref{sketch}A). Remarkably, the excitonic nature of the cavity's mirrors endows the system with two desirable features not present in conventional cavities: (1) a momentum-independent optical mode's energy, and (2) spin-polarized cavity modes that split due to the valley Zeeman effect under an external magnetic field. Moreover, we demonstrate the excitonic saturation of the optical mode as a function of pump power and show its tunability via electrical manipulation of each monolayer mirror. 

%%%%%%%%%%%%%%%%%%%%%%%%%%%%%%%%%%%%%%%%%%%%%%%%%%%%%%%%%%%%%%%%%%%%

\subsection{Device design and simulation}

The realization of this cavity based on atomically thin materials relies on the high reflection from monolayer MoSe$_2$ at the excitonic resonance. In particular, an optically thin material can act as a narrow-band resonant mirror near its optical resonance. For excitons in atomically thin materials, the reflectance reaches its peak at the exciton wavelength, with a value determined by the ratio $\Gamma_{\text{r}}/\Gamma_{\text{nr}}$ (see Supplementary Text). Exciton reflectances of more than 85$\%$ have been experimentally realized in MoSe$_2$, thanks to the substantial oscillator strengths and relatively low non-radiative rates of the excitons in TMDs when they are encapsulated inside hexagonal boron nitride (hBN) \cite{cadiz2017excitonic,Scuri2018} (Fig.~\ref{sketch}A). Utilizing such an effect, we can stack two monolayers vertically to form an optical cavity at the exciton wavelength with a thickness determined by the dielectric spacer (Fig.~\ref{sketch}A and B).

The demonstration of such a cavity, however, imposes substantial experimental challenges. Strains and disorders introduced during the assembly of van der Waals (vdW) heterostructures can not only enhance the non-radiative processes, which reduces the reflectance of the monolayer but also lead to inhomogeneous broadening and the variation of exciton energies across the samples. This makes it difficult to realize high reflectivity in both TMD monolayers while exactly matching the energies of the reflection peaks. To overcome this challenge, we first assemble a series of hBN-encapsulated monolayer MoSe$_2$ on a silicon substrate and characterize their optical response at 4 K. In doing so, we can select regions of two TMD samples with not only high exciton reflectivity but also similar exciton energy. We then transfer one heterostructure on top of the other to form the cavity. This offers a more controllable way of fabricating cavities as it involves only a single transfer step that will likely introduce less strain and disorders than the full assembly of the vdW structure.

\begin{figure*}
    \centering
    \includegraphics[width=1.3\columnwidth]{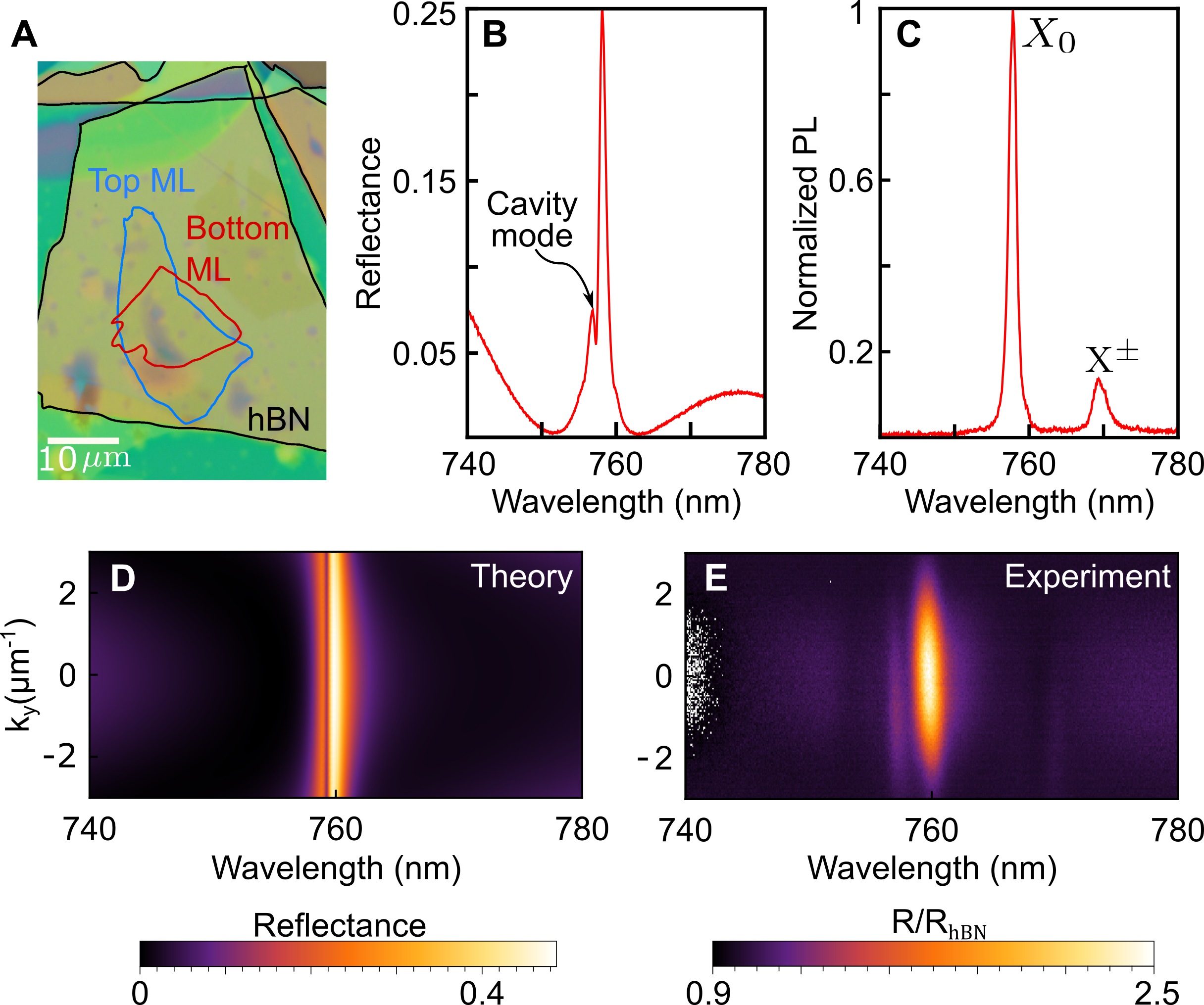}
    \caption{Experimental characterization of the cavity sample. a) Microscope picture of the nano-cavity device with hBN layers and MoSe$_2$ bottom and top layers indicated in black, red, and blue, respectively. b) Experimental reflectance spectrum of the device. As theoretically predicted, the confined optical mode manifests as a narrow minimum in the reflectance spectrum (indicated by the arrow). c) PL spectrum of the device. The central emission (X$_{0}$) coincides with the device's resonant wavelength. A secondary peak from charged excitonic states is observed at longer wavelengths (X$^{\pm}$). d) Nanocavity dispersion calculated via TMM simulation. The confinement mechanism makes the cavity mode flat within our NA. e) Experimental dispersion of the device measured via far-field imaging. As theoretically predicted, the optical mode is flat in momentum.}
    \label{device}
\end{figure*}

In our experiments, we design the device such that two monolayer mirrors of MoSe2 are embedded inside layers of hBN encapsulation with a total hBN thickness of 240 nm (Fig.~\ref{sketch}B). The monolayers are positioned symmetrically at 60 nm from the top and bottom of the vdW heterostructure and are individually contacted using the Si substrate as a back gate. This gives control over the exciton energies and decay rates (a schematic of the device is shown in Fig.~\ref{sketch}B. The thickness of various layers is chosen based on a transfer matrix method (TMM) simulation. Importantly, we note that the TMDs separation is slightly shorter than half the wavelength of the exciton resonance (in the hBN medium). At precisely the half-wavelength condition, the standing cavity mode becomes optically dark because of the vanishing electric field at the locations of the TMDs, and can only decay by non-radiative means. This would be analogous to a Fabry-Perot cavity with finite absorption loss and zero transmission mirrors (see Supplementary Text for additional discussion). Meanwhile, the total thickness of the hBN obeys the requirement of having minimum reflectance from the SiO$_2$ substrate-hBN system at the exciton resonance, which facilitates the identification of the optical mode (see Supplementary Text for a complete discussion about the role of the hBN thickness).

\begin{figure*}
    \centering
    \includegraphics[width=1.5\columnwidth]{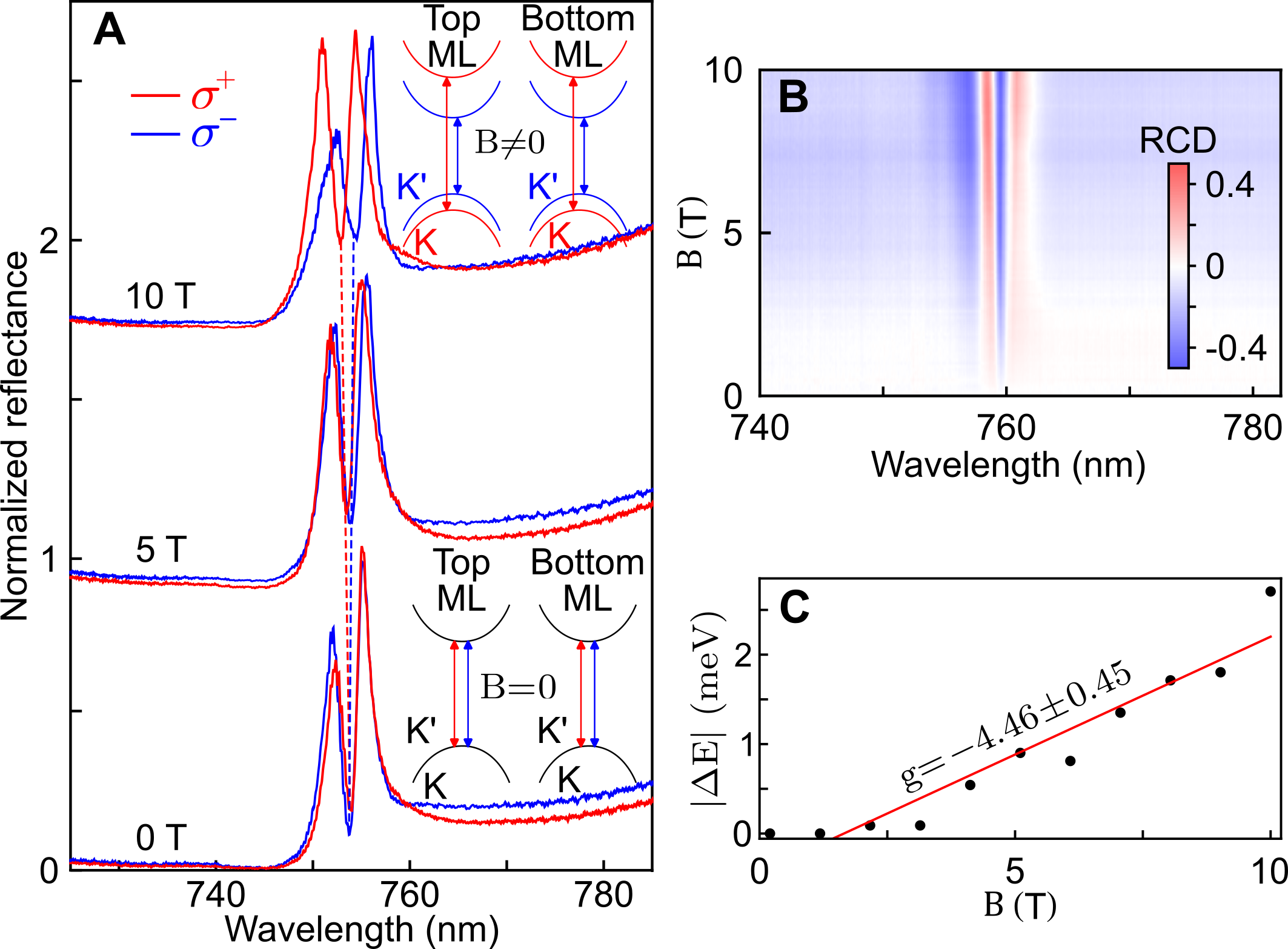}
    \caption{Chiral behavior induced by an external magnetic field $B$. a) Device's reflectance spectra for the orthogonal circular polarization states $\sigma^+$ (blue) and $\sigma^-$ (red) at three different values of $B$: 0 T (bottom), 5 T (middle), and 10 T (top). The insets show a depiction of the mechanism by which the chirality is established: the modes are degenerate in the absence of a magnetic field (black scheme), but the mode split in the presence of a magnetic field and exhibits a chiral light-matter response  (blue and red scheme). The data is collected at a different spot than figure \ref{device}. b) Reflective circular dichroism of the nano-cavity device for increasing $B$. c) Energy difference between the $\sigma^+$ and $\sigma^-$ cavity modes as a function of $B$. A linear regression indicates a value of the magnetic factor g$=\!-4.46\pm0.45$; in good agreement with reported values and theoretical predictions. }
    \label{magnet}
\end{figure*}

Figures \ref{sketch}C-D show the reflectance spectra of two MoSe$_2$ monolayers at 4 K, before they were stacked together to form the cavity. The reflectance spectra can be fitted using the TMM and modeling the 2D material's optical response as a Lorentz oscillator \cite{Yu2010,Glazov2014,Scuri2018,Back2018}, from which we extract $\Gamma_{\rm r}$ and $\Gamma_{\rm nr}$ (see Supplementary Text). With this information, we proceed to verify the device design using TMM and finite-difference time-domain (FDTD) simulations and extract reflectance and effective mode length (Fig.~\ref{sketch}E see Methods for further details). The effective mode length, displayed in the inset, is a measure of the distribution of the optical field, which is equal to the physical thickness of the device for non-resonant wavelengths, and becomes reduced if there is a localization of the electromagnetic field (see Supplementary Text for its mathematical definition and related discussion). The emergence of a minimum in both the reflectance and effective mode length (Fig.~\ref{sketch}E), as well as the electric field intensity distribution (Fig.~\ref{sketch}F) confirm the formation of a standing optical mode at a wavelength $\lambda\!\approx\!756$ nm. The enhancement of the electric field inside the cavity volume strongly depends on the substrate, as discussed in the Supplementary Text. FDTD simulations indicate that this optical mode has a quality factor of $Q\!\approx\!1060$ at the resonance wavelength. In the absence of non-radiative losses, the separation between TMDs can be arbitrarily close to $\lambda/2$, with a cavity mode of an infinitely long lifetime. However, non-radiative processes lead to the disappearance of such modes. Therefore, $\Gamma_{\text{nr}}$ sets the limit for how close to $\lambda/2$ the cavity length can be while still having a detectable cavity mode (Supplementary Text). 

Figure \ref{sketch}G shows TMM simulations of the reflectance spectrum of such a vdW heterostructure for variable energy detuning between the top (X$_{\rm top}$) and bottom (X$_{\rm bottom}$) monolayers' excitons, using the obtained values of $\Gamma_{\text r}$ and $\Gamma_{\text nr}$ for each monolayer. For detuned resonances, the spectrum shows two individual reflectance peaks, but as one approaches degeneracy, the optical mode establishes, as highlighted by the encircled area in the figure. Importantly, the simulation predicts a tunability of the cavity mode by changing the detuning between the component mirrors.

%%%%%%%%%%%%%%%%%%%%%%%%%%%%%%%%%%%%%%%%%%%%%%%%%%%%%%%%%%%%%%%%%%%%

\subsection{Nano-cavity device characterization}

A microscope picture of the final device is presented in Fig.~\ref{device}A. The edges of the hBN, top, and bottom MoSe$_2$ monolayers, are indicated in black, blue, and red, respectively. The experimental reflectance is presented in Fig.~\ref{device}B. The narrow dip in the reflectance, in agreement with the prediction in Fig.~\ref{sketch}E, confirms the presence of the confined electromagnetic mode in the nano-cavity. Fig.~\ref{device}C shows the photoluminescence (PL) spectrum of the sample at the same spot. Unlike reflectance spectra with complex lineshape, the exciton emission (X$_0$) is a Lorentzian peak centered at the cavity mode. In addition, a secondary peak from charged excitonic states (X$^{\pm}$) is detected at longer wavelengths, which is not resonant with the optical mode \cite{Shepard2017}. A particular feature of this architecture is shown in panels D and E of Fig.~\ref{device}: the cavity mode presents a flat momentum dispersion. This is verified both theoretically via TMM simulations (panel D) and experimentally by realizing Fourier spectroscopic measurements of the far-field of the cavity mode (panel E). In conventional planar cavities, varying angles lead to different optical path lengths and phases. In our design, the exciton energy is agnostic to varying angles of incidence and the propagation phase is canceled out by the wavelength-dependent phase of the Lorentz oscillator, making the cavity dispersion flat. Further discussion about the origin of this flat band can be found in the Supplementary Text.

%%%%%%%%%%%%%%%%%%%%%%%%%%%%%%%%%%%%%%%%%%%%%%%%%%%%%%%%%%%%%%%%%%%%

\subsection{Magnetically-induced chirality}

The excitonic origin of the high reflectance in the monolayer TMD mirrors endows the nano-cavity with another unique capability, i.e., a chiral behavior induced by an external magnetic field ($B$). A potential application of such a chiral cavity is the circular polarization-dependent light-matter interaction that can be achieved when embedding an unpolarized emitter resonant with one of the chiral modes. In this scenario, the emitter will be transparent to one of the cavity modes, but interact with the other one \cite{hallett2022engineering}. To demonstrate this chirality, we measure the $B$-dependent reflectance when illuminating the sample with opposite circular polarization ($\sigma^+$ and $\sigma^-$) in a Faraday configuration (see Methods). Fig.~\ref{magnet}A shows the reflectance spectra of the device for the two circular polarization states under three different values of $B$: 0 T, 5 T, and 10 T. We note that this measurement was performed in a different spot on the sample with respect to Fig.~\ref{device} resulting in slight spectral differences. The small difference between the $\sigma^+$ and $\sigma^-$ spectra at 0 T originates in the imperfect polarization filter. Regardless, the degeneracy of the mode is evident in the absence of a magnetic field. This degeneracy of the two optical modes with orthogonal circular polarization is lifted with increasing $B$. This behavior originates in the valley-dependent optical selection rule and the valley Zeeman effect of TMD monolayers due to inversion symmetry breaking and spin-orbit coupling\cite{Xiao2012,Xuan2020}: a magnetic field splits the energy of excitons in the K and -K valleys, which shift the reflectance peak and the cavity mode of $\sigma^+$ and $\sigma^-$ light (inset of Fig.~\ref{magnet}A). Such magnetic tuning of chiral light-matter coupling is usually not achievable in photonic microcavities due to the non-magnetic nature of the typical component materials \cite{Megahd2022}. Here, excitons in MoSe2 experience the Zeeman effect which introduces the chiral behavior.

To perform a quantitative analysis of the cavity's chiral behavior, we collect data over the range of 0 T to 10 T. The reflective circular dichroism (RCD), defined as $\frac{{\rm R}^{+}-\rm{R}^{-}}{{\rm R}^{+}+\rm{R}^{-}}$, where $R^{+(-)}$ is the reflectance of the polarization $\sigma^{+(-)}$, is shown in Fig.~\ref{magnet}B. The chiral behavior manifests as an increasing RCD with increasing magnetic field $B$, which reaches a value of 0.41 at the highest magnetic field of 10 T. By fitting the energy splitting of the two chiral modes in response to the magnetic field to the relationship $\Delta E\!=\!g \mu_B B$ (where $\mu_B$ is the Bohr magneton), we obtain g$=\! -4.46\pm0.45$. The extracted g-factor is in good agreement with previously reported values and theoretical predictions \cite{Srivastava2015,aivazian2015magnetic,robert2021measurement}. Measurements of the magnetically induced chirality performed on a third spot of the sample show consistent results (Supplementary Text).

%%%%%%%%%%%%%%%%%%%%%%%%%%%%%%%%%%%%%%%%%%%%%%%%%%%%%%%%%%%%%%%%%%%%

\begin{figure*}
    \centering
  \includegraphics[width=1.5\columnwidth]{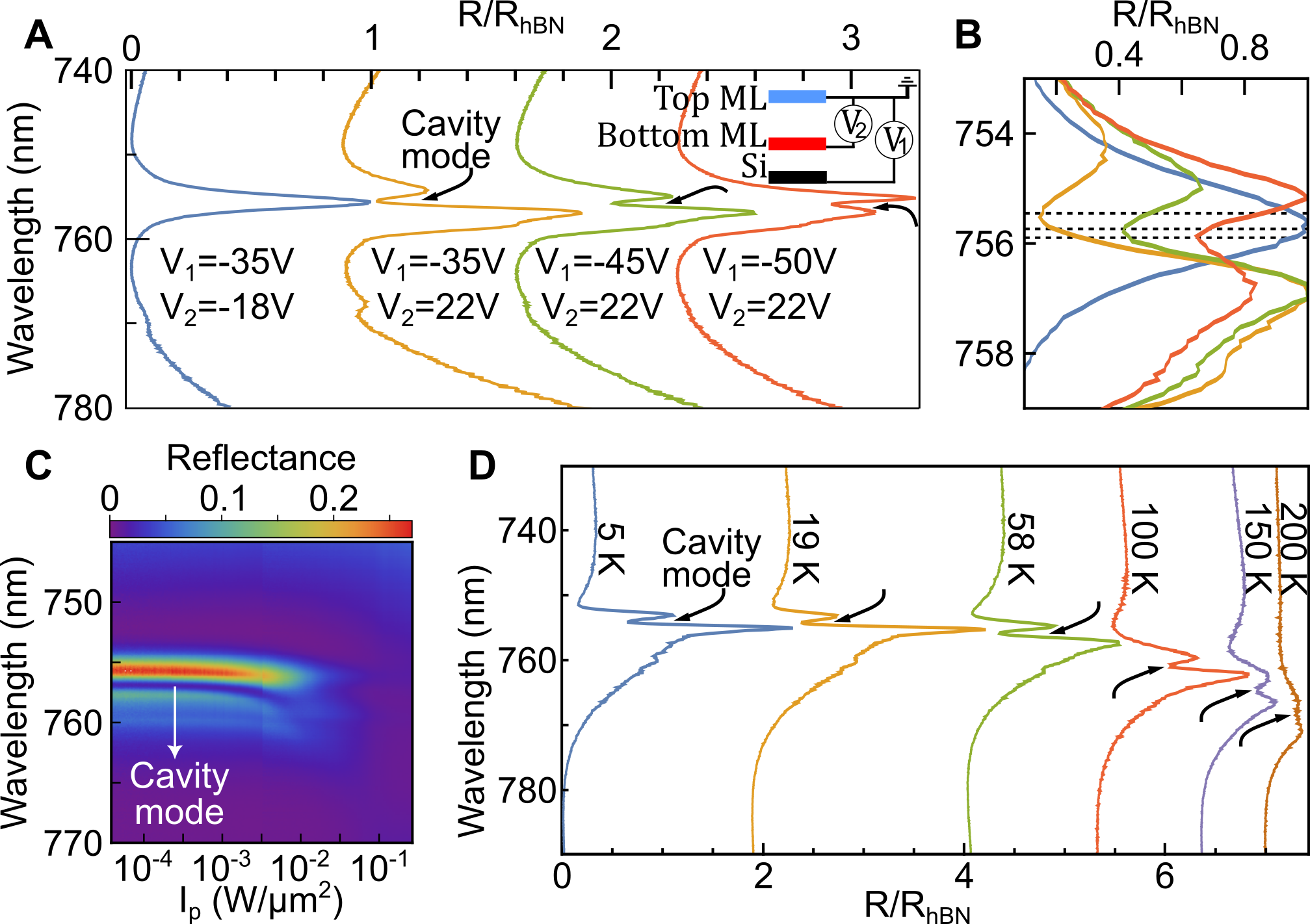}
    \caption{Tuning mechanisms of the optical mode. a) Independent electrical contacts on each MoSe$_2$ mirror (as shown in the inset) give control over the charge density, and hence over the excitonic resonance energy and oscillator strength. As a result, the cavity mode can be turned off (Blue line) or it can be tuned over a range of $\sim0.5$ nm, as shown in panel b. c) Pump intensity-dependent reflectance spectrum. The pump intensity axis (horizontal) does not follow a linear trend, because the power was not modified linearly. As the pump power increases, the optical mode red-shifts and broadens due to thermal fluctuations and saturation of the TMDs. After a critical intensity $I_p\!\approx\!7\!\times\!10^{-3}$ W$/\mu$m$^2$, the mode completely vanishes. d) Temperature dependence of the optical mode collected with a $10^{-4}$ W/$\mu$m$^2$ white pump. The data shows a tunability of $\approx\!10$ nm in the range of 4 K to 100 K. At $T\!=\!200$ K the mode is not identifiable anymore. In panels a) and d), the arrows serve as a guide for the eye to track the modification of the optical mode.}
    \label{power_temp}
\end{figure*}

\subsection{Electrical tuning and thermal response of the optical mode}

We now demonstrate some of the mechanisms that can be implemented to obtain control over the optical mode. In Fig.~\ref{power_temp}, we show that the resonance can be modified by electrical gating, and optical pumping. Each monolayer mirror is individually electrically contacted as detailed in the Methods section. As shown in Fig.~\ref{power_temp}A as a proof-of-principle, the electrical gating allows us to turn on and off the mode in real-time, as well as to tune its energy (panel B shows an inset of the resonance with continuous tunability of $\sim0.5$ nm). Further optimization of the gates can lead to greater tunability at lower voltages. The optical power also offers a control tool over the cavity resonance. Figure \ref{power_temp}C shows the reflectance spectra of the sample when excited by a supercontinuum white laser (with a pulse duration of $\approx$ 1 ns) at varying power levels. The cavity mode is not notably modified below $I_p\!\approx\!10^{-3}$ W$/\mu$m$^2$. Upon further increasing the laser intensity, the mode red-shifts and broadens. Finally, above a threshold pump intensity $I_p\!\approx\!7\!\times\!10^{-3}$ W$/\mu$m$^2$, the excitonic response and the cavity mode disappear in the reflectance spectra. This response to increasing pump powers could stem from the subsequent saturation of the exciton resonance and the laser-induced exciton decoherence (linewidth broadening and increasing $\Gamma_{\rm{nr}}$).

We therefore further investigate the thermal effects by measuring the reflectance spectra at different temperatures, as shown in Fig.~\ref{power_temp}D. We normalize the device's reflectance to that of a region without TMDs but with the same hBN thickness at each temperature. With increasing temperatures, the exciton and optical modes broaden and red-shift, in a similar fashion observed in Fig.~\ref{power_temp}C. The optical mode is robust up to a temperature $T\approx\!100$ K, while the temperature tunes cavity mode over a range of $\approx\!10$ nm with a mode broadening of a factor $\approx\!1.3$. The thermal fluctuations enhance the non-radiative decay rate of the exciton by promoting the exciton-phonon decoherence, reducing the ratio $\Gamma_{\rm r}/\Gamma_{\rm nr}$ to the critical point where the mode vanishes at $T\approx200$ K \cite{Jakubczyk_2018,Epstein_2020}.

%%%%%%%%%%%%%%%%%%%%%%%%%%%%%%%%%%%%%%%%%%%%%%%%%%%%%%%%%%%%%%%%%%%%

\subsection{Discussion and outlook}

We demonstrate a mechanism to achieve light confinement by harnessing the high quality of the optical excitations in TMD materials, using them as atomically thin mirrors. The 2D nature of the constituent mirrors endows this device with advantages in terms of miniaturization and integration capabilities with a Q-factor comparable to traditionally used planar cavities (see Table S1 in Supplementary Materials). Future improvements in the materials’ quality and heterostructure fabrication, such as nano-squeegee\cite{rosenberger2018nano} and laser annealing \cite{shree2019high,rogers2018laser}, could further reduce their optical loss by decreasing the non-radiative and inhomogeneous broadening associated with disorders and lead to large-area samples. For instance, TMD monolayers grown by chemical vapor deposition have been shown to exhibit optical quality comparable with the best hBN-encapsulated samples \cite{shree2019high,rogers2018laser}.

Remarkably, the demonstrated flat dispersion sets this architecture apart from conventional planar cavities, which have a strong angular dependence on the resonant mode energy. The weak dispersion of a TMD cavity can enable the efficient control of point emitters, for example, quantum emitters in 2D materials \cite{palacios2017large,azzam2021prospects,thureja2024electrically,srivastava2015optically}, without complicated photonic structures such as photonic crystals and curved mirrors. Furthermore, such a concept can be extended to other resonant optical effects in not only 2D materials (including other TMDs and hBN \cite{Ma_2022}) but also other optically thin systems, which could enable a wide range of optical applications covering visible and IR spectra. 

The unique properties of chiral cavities have been proposed for the realization of non-equilibrium states of matter and new topological effects \cite{Hubener2020,dag2023cavity,jiang2023engineering}. Other architectures have been demonstrated to host chiral optical modes \cite{rechcinska2019engineering}, and our design opens an alternative to exploring such configurations in a vdW heterostructure. There are also intriguing analogies between a TMD mirror and a dipole array of cold atoms \cite{rui2020subradiant} that could make this architecture suitable for the study of subradiant/superradiant states and other collective effects in a condensed matter setting \cite{Guimond2019}. The demonstrated electrical tuning of the cavity resonance enables a dynamic control for devices such as tunable optical isolators and polarization-dependent light-microwave transductors \cite{Zhou_2020, Gao_2020}. Furthermore, optically pumping of the valley polarization in TMDs opens up the intriguing possibility of ultrafast optical control of the chiral light-matter coupling \cite{Guddala_2021, liu2020generation, Sun_2017, Hao_2022, Wang_2018}. 
Finally, motivated by recent observations of strongly interacting excitons in hetero-bilayer TMDs superlattice structures \cite{xiong2023correlated,park2023dipole,gao2023excitonic}, one can envisage embedding such lattices inside our planar cavity and explore the rich physics of Bose-Hubbard polaritonic models \cite{angelakis2017quantum,hartmann2016quantum}.

%%%%%%%%%%%%%%%% MATERIALS AND METHODS %%%%%%%%%%%%%%%

\subsection{Materials and Methods}

\textbf{TMM and FDTD numerical simulations:} For the numerical simulations of the device's reflectance, we apply the transfer matrix method (TMM), by using the refractive indices of the different materials from a database. In this formalism, we simulate the response of the system to an incoming plane wave at any angle, which in this case was chosen to be $0^{\circ}$ (except for Fig.~\ref{device}C, where we calculate the angular dependence). For the calculation of the electric field intensity profile, effective cavity mode length, and Q factor, we rely on a finite-difference time-domain (FDTD) simulation implemented in commercial software. We simulate the situations of the cavity excited through an embedded dipole source, or by an incident plane wave.

\textbf{Device fabrication:} Monolayer MoSe$_2$ and hBN flakes are exfoliated from bulk crystals on top of silicon substrates with an oxide layer of 285 nm. High optical quality monolayers of MoSe$_2$ are obtained using the QPress Exfoliator at CFN in Brookhaven National Lab. The MoSe$_2$ flakes are identified under an optical microscope and confirmed by photoluminescence measurements. The thickness of hBN flakes is verified by atomic force microscopy. The device is assembled by the van der Waals (vdW) dry transfer technique. We first fabricate several hBN/MoSe$_2$/hBN heterostructures and characterize them by photoluminescence and reflectance measurements to find the ideal spot. Finally, to obtain the nano-cavity device, two heterostructures with identical excitonic energies and high radiative decay rates are carefully chosen and stacked together.

\textbf{Fabrication of electrical contacts:} To expose and contact the MoSe$_2$ monolayers respectively, we first define the area of etching using electron-beam lithography and etch the hBN using inductively coupled plasma (ICP) reactive ion etcher. Then we make the electrical contacts to both MoSe$_2$ monolayers using chromium (5nm) and gold (90nm) deposited via thermal evaporation to connect them to the wire-bonding pads.

\textbf{Setup for optical measurements:} The sample is held at a temperature of 6 K in a closed-loop cryostat. For the optical measurements, we use a confocal microscopy setup in a reflectance configuration. The focused laser spot on the sample is about 1 $\mu m$ in diameter. A tungsten lamp and a supercontinuum white laser are employed as broadband light sources; there is no difference in the results obtained with each source. For photoluminescence measurements, we use a HeNe laser ($\approx633$ nm) to excite the material. For the measurements of the magnetic field dependence, polarization-resolved reflectance spectra were collected by placing a quarter-wave plate followed by a linear polarizer in the detection path. The signal is finally collected by a charge-coupled device attached to a spectrometer. A detailed description of this setup can be found in Ref.~\cite{Sell2022}.

%%%%%%%%%%%%%%%%%%%%%%%%%%%%%%%%%%%%%%%%%%%%%%%%%%%%%%%%%%%%%%%%

\subsection{Acknowledgements}

The authors thank Alejandro Gonz\'alez-Tudela, Edo Waks, Xavier Marie, and Dominik Wild for valuable discussions and enriching feedback on the manuscript.

\textbf{Funding:}
This work was supported by the NSF DMR-2145712, AFOSR FA9550-19-1-0399, FA9550-22-1-0339 and ONR N00014-20-1-2325, NSF IMOD DMR-2019444, ARL W911NF1920181, Minta Martin and Simons Foundation. The sample fabrication was supported by the U.S. Department of Energy, Office of Science, Office of Basic Energy Sciences Early Career Research Program under Award No. DE-SC-0022885. K.W. and T.T. acknowledge support from the JSPS KAKENHI (Grant Numbers 20H00354, 21H05233 and 23H02052) and World Premier International Research Center Initiative (WPI), MEXT, Japan. This research used Quantum Material Press (QPress) of the Center for Functional Nanomaterials (CFN), which is a U.S. Department of Energy Office of Science User Facility, at Brookhaven National Laboratory under Contract No. DE-SC0012704.

\textbf{Author contributions:} 
D.G.S.F., S.S., M.J.M., R.N., Y.Z., and M.H. conceived and designed the experiments. S.S., D.G.S.F., and M.J.M. performed the simulations. K.W., T.T., S.P., and H.J. supplied the necessary material for the fabrication of the sample. R.N. fabricated the samples. D.G.S.F. and S.S. performed the experiments with assistance from E.M. and V.S. D.G.S.F. and S.S. analyzed the data and interpreted the results with help from M.J.M. and A.G. D.G.S.F. and S.S. wrote the manuscript, with input from all authors. All work was supervised by M.H. and Y.Z.

\textbf{Competing interests:}
There are no competing interests to declare.

\textbf{Data and materials availability:}
All data needed to evaluate the conclusions in the paper are present in the figures of the paper and/or the Supplementary Materials. The data can also be accessed at Zenodo \cite{dataset}.\\

\small{
\noindent$^{\dagger}$hafezi@umd.edu\\
$^{\ddag}$youzhou@umd.edu}

\bibliography{biblio}

\newpage
%%%%%%%%%% Merge with supplemental materials %%%%%%%%%%
%%%%%%%%%% Prefix a "S" to all equations, figures, and tables and reset the counter %%%%%%%%%%

\setcounter{equation}{0}
\setcounter{figure}{0}
\setcounter{table}{0}
\setcounter{page}{1}
\makeatletter
\renewcommand{\theequation}{S\arabic{equation}}
\renewcommand{\thefigure}{S\arabic{figure}}
\pagenumbering{roman}

\subsection{\Large Supplementary Information}

%%%%%%%%%%%%%%%%%%%%%%%%%%%%%%%%%%%%%%%%%%%%%%%%%%%%%%%%%%%%%%%%%%%%%%%%%%%%%%%%%%%%%%%%%%%%%%%%%%%%%%%%%%%%%%%%%%%%%%%%%%%%%%

\subsection{Contents}

\noindent {\bf \hyperref[si:1]{1. Device fabrication}}
\\{\bf \hyperref[si:2]{2. A $\lambda_0/2$ cavity}}
\\{\bf \hyperref[si:3]{3. Role of the \lowercase{h}BN thickness}}
\\{\bf \hyperref[si:4]{4. Interplay between $\Gamma_{\text r}$ and $\Gamma_{\text{nr}}$}}
\\{\bf \hyperref[si:5]{5. Simulation of different cavity configurations}}
\\{\bf \hyperref[si:6]{6. Quality factor and effective mode length}}
\\{\bf \hyperref[si:7]{7. Optical mode's angular dependence}}
\\{\bf\hyperref[si:8]{8. Identification of cavity mode vs non-interacting oscillators}}
\\{\bf \hyperref[si:9]{9. Additional data: magnetic field-induced chirality}}
\\{\bf \hyperref[tab:Q]{Table I. Q-factors of planar cavities reported in the literature.}}

\newpage
%%%%%%%%%%%%%%%%%%%%%%%%%%%%%%%%%%%%%%%%%%%%%%%%%%%%%%%%%%%%%%%%%%%%%%%%%%%%%%%%%%%%%%%%%%%%%%%%%%%%%%%%%%%%%%%%%%%%%%%%%%%%%%

\subsection{\label{si:1} 1. Modeling a TMD monolayer as a Lorentz oscillator}

The strong modulation of the refractive index of a TMD monolayer upon driving the excitonic resonance can be well described by a Lorentz oscillator model. By fitting the reflectance spectrum of each monolayer-hBN heterostructure (before stacking) to such a model, using a Transfer Matrix Method (TMM) formalism (as described in the Methods section), we extract the real and imaginary parts of the refractive indices of the top and bottom monolayers. We also use the decay parameters of the TMD mirror reported by Scuri \textit{et al.} \cite{Scuri2018} to estimate the performance of a device made with two such monolayers. We use their parameters as they reported a monolayer with the best optical quality and highest reflectance. This anticipates what one can expect from a TMD cavity made of monolayers of superior optical quality. In this theoretical formalism, the susceptibility of the TMD monolayer is given by 
\begin{equation}
    \chi (\omega) = -\dfrac{c}{\omega_0 d} \dfrac{\Gamma_r}{(\omega - \omega_0)+i\Gamma_{nr}/2},
\end{equation}
\noindent where $\omega_0$ is the excitonic resonance frequency, $d$ ($= 0.7$ nm) is the thickness of a TMD monolayer, and $\Gamma_r$ and $\Gamma_{nr}$ are the radiative and non-radiative decay rates, respectively. $\Gamma_r$ is the rate at which an exciton in the monolayer recombines by emitting a photon, whereas, $\Gamma_{nr}$ is the non-radiative recombination rate via processes such as phonon scattering. The frequency-dependent complex refractive index of the TMD monolayer is then
\begin{equation}
    n_{\text{TMD}} (\omega) = \sqrt{n_{\text{bulk}}^2 (\omega) - \dfrac{c}{\omega_0 d} \dfrac{\Gamma_r}{(\omega - \omega_0)+i\Gamma_{nr}/2}},
\end{equation}
\noindent where $n_{\text{bulk}}$ is the refractive index of the bulk TMD material (and hence without the excitonic resonance).

From the fittings (as shown in Fig. 1 in the main text) we obtain the following values:\\
Top monolayer: $\Gamma_r=2.45$ meV, $\Gamma_{nr}=0.5$ meV, $\lambda_0=756.2$ nm.\\
Bottom monolayer: $\Gamma_r=2.15$ meV, $\Gamma_{nr}=1$ meV, $\lambda_0=756.2$ nm.\\
For comparison, the decay rates of the monolayer reported in Scuri \textit{et al.} \cite{Scuri2018}: $\Gamma_r=4.38$ meV, $\Gamma_{nr}=0.2$ meV.

Although the resonant wavelength of the monolayer TMD mirror demonstrated in Scuri \textit{et al.} \cite{Scuri2018} differs from our two monolayers, for theoretical comparison purposes, we assume it has the same resonance wavelength, i.e., $\lambda_0=756.2$ nm. In the following sections, we use the obtained refractive indices of the aforementioned three monolayers to study different cavity geometries. For the TMM and FDTD simulations, we use the refractive indices obtained from Ref.~\cite{Palik}, which at the exciton resonance wavelength correspond to n$_{\text{hBN}}(756.2\rm{\ nm})\approx2.1$, n$_{\text{SiO}_2}(756.2\rm{\ nm})\approx1.5$, n$_{\text{Si}}(756.2\rm{\ nm})\approx3.7$.

%%%%%%%%%%%%%%%%%%%%%%%%%%%%%%%%%%%%%%%%%%%%%%%%%%%%%%%%%%%%%%%%%%%%%%%%%%%%%%%%%%%%%%%%%%%%%%%%%%%%%%%%%%%%%%%%%%%%%%%%%%%%%%

\subsection{\label{si:2} 2. A $\lambda_0/2$ cavity}

\begin{figure}[ht]
    \centering
    \includegraphics[width=\columnwidth]{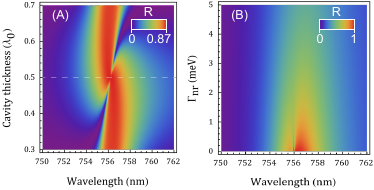}
    \caption{\textbf{Reflectance of the cavity as a function of cavity thickness and non-radiative decay rate.} (A) Reflectance of the cavity formed by two TMD monolayers ($\lambda_0=756.2$ nm, $\Gamma_{\text{r}}=4.38$ meV, $\Gamma_{\text{nr}}=0.2$ meV) in air for different cavity thickness. The mode is not visible in the reflectance spectrum when the thickness of the cavity is exactly $\lambda_0/2$, as marked by the white dashed line. (B) Reflectance of the cavity for varying $\Gamma_{\text{nr}}$ and a cavity thickness of 350 nm ($\sim 0.46 \lambda_0$).}
    \label{lambdahalf}
\end{figure}

In the most simplistic scenario, one can imagine a cavity formed by two TMD monolayers separated by air. From the physics of Fabry-Perot cavities, we intuitively expect to have the best optical confinement when the spacing between the TMD monolayers is an integer factor of half of the exciton resonance wavelength $\lambda_0$ because at this energy the reflectivity from the TMD is maximized. In Fig. \ref{lambdahalf}A, we simulate the spectral reflectance $R$ for different cavity thicknesses. For our simulations, we consider the excitonic resonance measured for the individual monolayers ($\lambda_0\!=756.2 \rm{\ nm}$), and decay constants $\Gamma_{\text{r}}=4.38$ meV, and $\Gamma_{\text{nr}}=0.2$ meV. The figure shows an intriguing dependence of the confined mode for variable cavity thickness: the linewidth reduces as one approaches the condition of $\lambda_0/2\approx 378.1$ nm (marked with a white dashed line). Generally, the energy of the mode is set by the TMD separation, while its quality factor is determined by $\Gamma_{\text{nr}}$ and the reflectivity (which is maximum at the exciton resonance). Interestingly, as one approaches the $n\lambda/2$ condition (with $n\in \mathbb{N}$), the mode becomes optically inaccessible. This can be understood in terms of the mode's radiative lifetime: for a TMD cavity with mirrors separated by $n\lambda/2$, the radiative lifetime is maximum and hence, a small $\Gamma_{\text{nr}}$ is enough to completely dampen the confined light. Therefore, this dark cavity mode is completely decoupled from the external electromagnetic fields, making it impossible to probe with an external source \cite{Guimond2019}.\\

In this context, the disappearance of the cavity mode from the reflectance spectrum is expected to be independent of $\Gamma_{\text{nr}}$ when the thickness of the cavity is exactly $\lambda_0/2$, as observed in Fig. \ref{lambdahalf}B. This is similar to an uncommon Fabry-Perot cavity with non-zero absorption and zero transmission. For our system, this particular behavior originates in the excitonic nature of the mirrors. For this reason, regardless of the non-radiative losses in the material, the mode will be optically dark at the $\lambda_0/2$ separation. This analysis was used to engineer the cavity thickness, making it slightly different from $\lambda_0/2$ to probe the cavity mode externally, as shown in the main text.

%%%%%%%%%%%%%%%%%%%%%%%%%%%%%%%%%%%%%%%%%%%%%%%%%%%%%%%%%%%%%%%%%%%%%%%%%%%%%%%%%%%%%%%%%%%%%%%%%%%%%%%%%%%%%%%%%%%%%%%%%%%%%%

\begin{figure*}
    \centering
    \includegraphics[width=1.5\columnwidth]{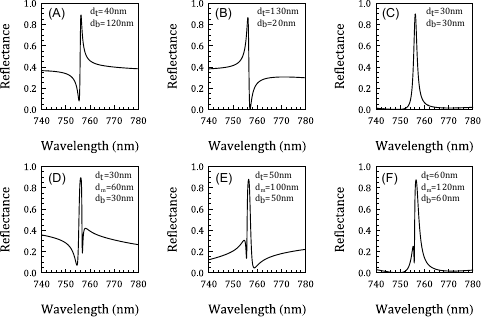}
    \caption{\textbf{Effect of the thickness of the encapsulating hBN on the reflectance of a TMD monolayer and the cavity heterostructure.} (A-C): Effect of the thickness of the encapsulating hBN on the reflectance of a TMD monolayer. The thicknesses of the (top, bottom) layer of hBN encapsulations are (A) (40 nm, 120 nm), (B) (130 nm, 20 nm), and (C) (30 nm, 30 nm). Different thicknesses of hBN lead to different reflectance spectra for the same monolayer. (D-F): Effect of the thickness of hBN on the reflectance of the cavity heterostructure. The thickness of the (top, middle, bottom) layers of hBN used for the simulations are (D) (30 nm, 60 nm, 30 nm), (E) (50 nm, 100 nm, 50 nm), and (F) (60 nm, 120 nm, 60 nm). The best cavity mode is obtained for the parameters of panel F, which are the nominal thicknesses of the final fabricated device.}
    \label{hbn}
\end{figure*}

\begin{figure*}[ht]
    \centering
    \includegraphics[width=1.5\columnwidth]{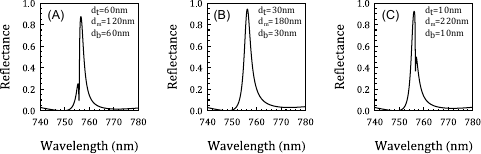}
    \caption{\textbf{Effect of the distribution of thickness of hBN on the detection of the cavity mode in reflection.} The thicknesses of the (top, middle, and bottom) layers of hBN used for the simulations are (A) (60 nm, 120 nm, 60 nm), (B) (30 nm, 180 nm, 30 nm), and (C) (10 nm, 220 nm, 10 nm) such that the total thickness of hBN is always 240 nm. When the MLs are separated a distance $d_m\!=\!\lambda/2n_{\text{hBN}}\!\approx\!180$ nm, the cavity mode is decoupled from the external field, making its detection impossible.}
    \label{hbn2}
\end{figure*}

\subsection{\label{si:3} 3. Role of the \lowercase{h}BN thickness}

The thickness of hBN plays a crucial role in the reflectance spectrum of both a TMD monolayer and the cavity heterostructure due to the interference of light from TMD and the background. To investigate the effect of hBN thickness on monolayer devices, we simulate three scenarios with different thicknesses of the top ($d_t$) and bottom ($d_b$) hBN encapsulations in an hBN/TMD/hBN heterostructure stacked on a Si substrate with 285 nm SiO$_2$, as used in experiments. In our simulations, we consider an excitonic resonance at 756.2 nm, and decay constants $\Gamma_{\text{r}}=4.38$ meV, and $\Gamma_{\text{nr}}=0.2$ meV. As shown in Fig. \ref{hbn}(A-C), depending on the thickness of hBN, the sample reflectance spectra can have different Fano-lineshapes or a reflectance peak. For the stacking of the cavity device, we chose the hBN thickness such that the exciton resonance manifests as a simple reflectance peak to avoid the complex lineshape and facilitate the identification and characterization of the optical confined mode.

\begin{figure*}[ht]
    \centering
    \includegraphics[width=1.5\columnwidth]{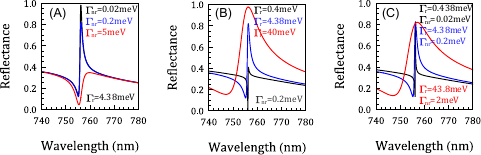}
    \caption{\textbf{Interplay between $\Gamma_{\text r}$ and $\Gamma_{\text{nr}}$.} (A) Reflectance of a monolayer with constant $\Gamma_{\text{r}}\!=\!4.38$ meV, and $\Gamma_{\text{nr}}\!=\!0.02$ meV (black), 0.2 meV (blue), and 5 meV (red). (B) Reflectance of a monolayer with constant $\Gamma_{\text{nr}}=0.2$ meV, and $\Gamma_{\text{r}}\!=\!0.4$ meV (black), 4.38 meV (blue), and 40 meV (red). (C) Reflectance of a monolayer with constant ratio $\Gamma_{\text{r}}/\Gamma_{\text{nr}}$. $(\Gamma_{\text{r}},\Gamma_{\text{nr}})=$ (0.438 meV, 0.02 meV) (black), (4.38 meV, 0.2 meV) (blue), (43.8 meV, 2 meV) (red). Both the top and bottom hBN encapsulations are chosen to be 60 nm thick for these simulations.}
    \label{gamma}
\end{figure*}

Next, we simulate the effect of hBN thickness on the reflectance spectrum of a cavity formed by stacking two monolayers on a similar substrate. The results of the simulations are shown in Fig. \ref{hbn}(D-F). Once again, we observe that different thicknesses of the top ($d_t$), middle ($d_m$), and bottom ($d_b$) hBN can drastically alter the reflectance spectrum of the cavity. The calculated reflectance spectra of samples with our designed device structures have minimal reflectance from the hBN around the cavity resonance frequency, as shown in panel F, making the cavity mode clearly identifiable.

From Fig.~\ref{hbn} we can conclude that the total thickness of hBN in the heterostructure, i.e., $d_t+d_m+d_b$, dictates the shape of the reflection spectrum. However, the distribution of the thickness of hBN is also crucial for the observation of the cavity mode. To see that, we fix the total thickness of hBN to 240 nm, and simulate different scenarios, as shown in Fig.~\ref{hbn2}. Interestingly, one can always observe the cavity mode in reflectance except in the scenario when the spacing between the two TMD monolayers is exactly half of the wavelength of the excitonic resonance in hBN, i.e., $d_m\!=\!\lambda_0/2n_{\text{hBN}}\approx180$ nm, where $n_{\text{hBN}} \approx 2.1$ is the refractive index of hBN at the resonance frequency. As discussed above, this happens because the electric field intensity of the cavity mode is zero at the position of the TMDs when $d_m$ is exactly equal to $\lambda_0/2n$.

%%%%%%%%%%%%%%%%%%%%%%%%%%%%%%%%%%%%%%%%%%%%%%%%%%%%%%%%%%%%%%%%%%%%%%%%%%%%%%%%%%%%%%%%%%%%%%%%%%%%%%%%%%%%%%%%%%%%%%%%%%%%%%

\subsection{\label{si:4} 4. Interplay between $\Gamma_{\text r}$ and $\Gamma_{\text{nr}}$}

The radiative ($\Gamma_{\text{r}}$) and non-radiative ($\Gamma_{\text{nr}}$) decay rates determine the peak reflectance of a monolayer, critically affecting its spectrum. Fig.~\ref{gamma} shows how $\Gamma_{\text{r}}$ and $\Gamma_{\text{nr}}$ determine the absolute reflectance of the monolayer, i.e., the quality of the mirror. In panel A, we use a fixed radiative decay rate of $\Gamma_{\text{r}}=4.38$ meV and vary the non-radiative decay rate. We observe that the monolayer acts as a better mirror if it has lower values of $\Gamma_{\text{nr}}$. At high values of $\Gamma_{\text{nr}}$, we only observe the absorption dip, and the reflectance peak almost disappears. Intuitively, with increasing non-radiative rates, the amount of light lost via non-radiative processes increases, which decreases the sample reflectance. This leads to the disappearance of the reflectance peak at higher values of $\Gamma_{\text{nr}}$.

Panel B shows the case where the value of $\Gamma_{\text{nr}}$ is fixed to 0.2 eV and $\Gamma_{\text{r}}$ varies. Once again, we observe that with increasing $\Gamma_{\text{r}}$ the peak reflectance increases and broadens. This can be understood as a reduction of the exciton lifetime (but the recombination is always radiative). 

Finally in panel C, both $\Gamma_{\text{r}}$ and $\Gamma_{\text{nr}}$ are varied, but the ratio $\Gamma_{\text{r}}/\Gamma_{\text{nr}}$ is kept constant. In this case, the maximum and minimum value of reflectance of the TMD monolayer is constant and only the linewidth of the reflectance peak is modified.

From all three scenarios in Fig.~\ref{gamma}, we also observe that increasing either of the decay rates also increases the linewidth of the reflectance spectrum as the linewidth of an emitter scales proportionally with the total decay rate. In summary, the higher $\Gamma_{\text{r}}$ is and the lower $\Gamma_{\text{nr}}$ is, the more similar the ML mirror is to an ideal perfect mirror.

%%%%%%%%%%%%%%%%%%%%%%%%%%%%%%%%%%%%%%%%%%%%%%%%%%%%%%%%%%%%%%%%%%%

\begin{figure*}[ht]
    \centering
    \includegraphics[width=1.5\columnwidth]{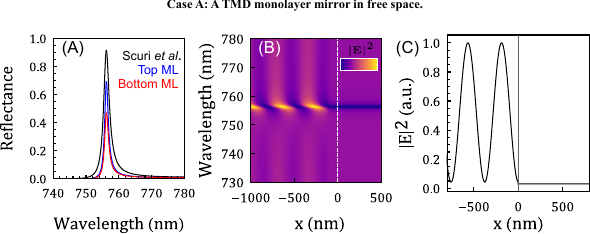}
    \caption{\textbf{A TMD monolayer mirror in free space.} (A) Simulated reflectance spectrum of the top (blue), bottom (red), and Scuri \textit{et al.} \cite{Scuri2018} (black) TMD monolayer (ML) mirror in free space. As expected from a two-level emitter, the incident light is reflected back at the excitonic resonance frequency. (B) Spatial electric field intensity profile $|E|^2$ as a function of wavelength when the TMD (top), located at $x=0$ nm as marked by a dashed vertical line, is illuminated with a plane wave source from the left. (C) Spatial intensity profile at the excitonic resonance frequency. On the left-hand side of the TMD (top), one can observe standing waves created by the incident and reflected light. The intensity drops on the right-hand side as the TMD reflects the incident light at the resonant frequency.}
    \label{1}
\end{figure*}

\begin{figure*}[ht]
    \centering
    \includegraphics[width=1.5\columnwidth]{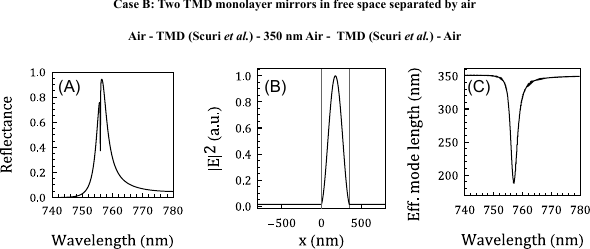}
    \caption{\textbf{Two TMD monolayer mirrors in free space separated by air.} (A) Simulated reflectance spectrum when two TMDs (Scuri \textit{et al.} \cite{Scuri2018}) are separated by 350 nm of air. The dip in the reflectance spectrum at $\lambda \sim 756$ nm indicates the formation of a cavity mode. (B) Spatial intensity profile at the cavity resonance frequency. The TMDs positions are marked by vertical lines. (C) Effective cavity mode length as a function of wavelength. Here the integration is performed over the TMD-air-TMD region. The estimated Q-factor of the cavity is $\sim 2828$.}
    \label{2}
\end{figure*}

\begin{figure*}[ht]
    \centering
    \includegraphics[width=1.5\columnwidth]{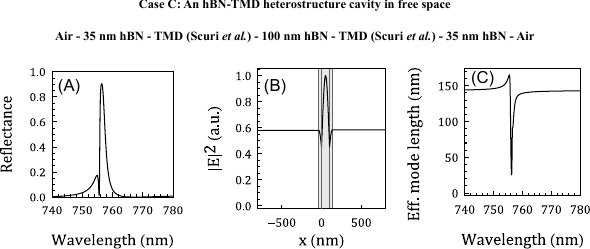}
    \caption{\textbf{An hBN-TMD heterostructure cavity in free space.} (A) Simulated reflectance spectrum for two TMDs (Scuri \textit{et al.} \cite{Scuri2018}) separated by 100 nm of hBN, and encapsulated by 35 nm of hBN on each side. (B) Spatial intensity profile at the cavity resonance frequency. The shaded region represents the hBN-TMD heterostructure cavity. (C) Effective cavity mode length as a function of wavelength. The estimated Q-factor of the cavity is $\sim 1219$.}% \pm 55$.}
    \label{3}
\end{figure*}

\begin{figure*}[ht]
    \centering
    \includegraphics[width=1.5\columnwidth]{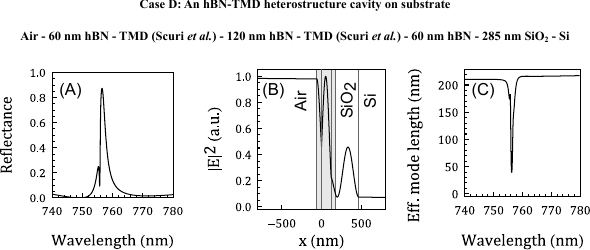}
    \caption{\textbf{An hBN-TMD heterostructure cavity on substrate.} (A) Simulated reflectance spectrum of the TMD-hBN heterostructure cavity on a SiO$_2$-Si substrate. (B) Spatial intensity profile at the cavity resonance frequency. The shaded region represents the hBN-TMD heterostructure cavity. (C) Effective cavity mode length as a function of wavelength. The estimated Q-factor of the cavity is $\sim 1308$.}% \pm 64$.}
    \label{4}
\end{figure*}

\subsection{\label{si:5} 5. Simulation of different cavity configurations}

We use TMM and FDTD techniques to simulate different cavity configurations. We start by simulating the reflectance spectrum and the spatial intensity profile when a TMD monolayer mirror is illuminated by a plane wave excitation (Fig.~\ref{1}). We then simulate the case where two TMD monolayer mirrors are separated by air (Fig.~\ref{2}). We also calculate the Q-factor and effective mode length for the resulting cavity. After that, we simulate the case of an hBN-TMD heterostructure consisting of two TMD monolayers separated by hBN in the middle and encapsulated by hBN on the top and bottom (Fig.~\ref{3}). Finally, we simulate the realistic scenario when such a heterostructure is placed on a Si substrate with a 285 nm thick layer of SiO$_2$ on top (Fig.~\ref{4}). 

The reflectance spectra of the devices are calculated by TMM and the 1D FDTD method, by illuminating the device with a broadband plane wave source and monitoring the reflected light. The cavity field profile, effective mode length, and Q-factor are simulated when exciting the cavity mode with a broadband dipole through FDTD simulations. For effective mode length calculations, we used the definition 

\begin{equation}
    L_{\text{eff}} = \dfrac{\left( \int \lvert E \rvert ^2 dx \right)^2}{\int \lvert E \rvert ^4 dx},
\end{equation}

\noindent where the integration is performed over the hBN/TMD/hBN/TMD/hBN heterostructure. This quantity is then a measure of the distribution of the electric field intensity in the nano-cavity. It is equal to the physical thickness of the device for non-resonant wavelengths (constant intensity along the full structure) and becomes reduced if there is a localization of the electromagnetic field.

Comparing panel B of figures \ref{2} to \ref{4}, one can notice the important role of the substrate in determining the enhancement of the electric field inside the cavity volume. In the three cases, the simulated MoSe$_2$ monolayers have the same $\Gamma_{\text{r}}$ and $\Gamma_{\text{nr}}$, but as it can be observed, the enhancement of the electric field is very different. Although we used the typical SiO$_2$ on Si substrate for this work, perspective future devices should consider a preliminary engineering of the substrate to maximize the enhancement of the electric field in the cavity.

%%%%%%%%%%%%%%%%%%%%%%%%%%%%%%%%%%%%%%%%%%%%%%%%%%%%%%%%%%%%%%%%%%%%%%%%%%

\subsection{\label{si:6} 6. Quality factor and effective mode length}

The quality factor and mode length of the cavity are important parameters, as they control the coupling efficiency ($g$) of an emitter inside the cavity as directly given by
\begin{equation}
    g = - \hat{e} \cdot \vec{d} \sqrt{\dfrac{\omega}{2\hbar \epsilon_0 V_{\rm{mode}}}} \sim \sqrt{\dfrac{1}{V_{\rm{mode}}}},
\end{equation}
\noindent where $\hat{e}$ is the polarization unit vector of the cavity mode at the position of the emitter, $\vec{d}$ is the dipole moment of the emitter, $V_{\rm{mode}} = L_{\rm{eff}}A$ is the mode volume for a spot of area $A$. For our experiments with a diffraction-limited spot, the mode volume is simply $V_{\rm{mode}} = L_{\rm{eff}} \times \left( \pi \mu \rm{m}^2 \right)$. Importantly, this tells us that a smaller mode volume, i.e., a smaller effective cavity length $L_{\rm{eff}}$ would result in stronger coupling to the emitter.
Moreover, the cavity can also modify the emission characteristics of the coupled emitter by enhancing the radiative decay rate by the Purcell factor $F_P$ which is given by
\begin{equation}
    F_P = \dfrac{3}{4\pi^2} \left( \dfrac{\lambda}{n} \right)^3 \dfrac{Q}{V_{\rm{mode}}} \sim \dfrac{Q}{V_{\rm{mode}}},
\end{equation}
\noindent which tells us that not only a smaller mode length, but a high Q-factor are desirable qualities for any cavity. The Q-factor of the cavity is given by the resonance frequency of the cavity, divided by the linewidth of the resonance, i.e., $Q=f_0/\Delta f$. Through FDTD simulations we get a quality factor of $Q\sim1060$ (which closely matches our reflectance data), and a mode length of $L_{\rm{eff}}\sim120$ nm. The $Q$ values can be improved further with better sample design and fabrication capabilities. Table~\ref{tab:Q} provides a comparison with planar distributed Bragg reflector (DBR) and metallic mirror cavities reported in the literature.

%%%%%%%%%%%%%%%%%%%%%%%%%%%%%%%%%%%%%%%%%%%%%%%%%%%%%%%%%%%%%%%%%%%%%%%%%

\subsection{\label{si:7} 7. Optical mode's angular dependence}

\begin{figure}[ht]
    \centering
    \includegraphics[width=\columnwidth]{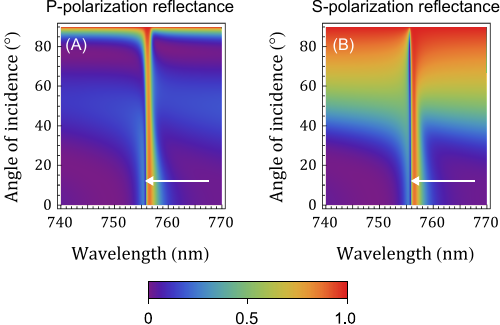}
    \caption{\textbf{Angular dependence of reflectance.} Calculated reflectance for (A) P and (B) S polarized light as a function of the angle of incidence for the configuration of Fig.~\ref{4}. The white arrows point out the mode for visual guidance. Unlike typical planar cavities, the energy of the cavity mode in our device does not change substantially with the angle of incidence. This is a consequence of the excitonic nature of the atomically-thin constituent mirrors. }
    \label{angle}
\end{figure}

In typical planar cavities, the photonic mode dispersion has a characteristic parabolic dependence. It originates in the different phase shifts that the light acquires as it propagates at different angles inside the cavity. Since the phase shift upon a reflection in a DBR or metallic mirror does not have any wavelength dependence, the acquired phase upon a cavity round-trip depends exclusively on the angle of incidence. A notable difference of the architecture that we present in this work, comes from the fact that the phase has a strong dependence on the light's wavelength near the resonance of a Lorentz oscillator. A variation in the propagation phase is compensated by a phase shift induced by a small change in the wavelength. An immediate consequence is the weak dispersion of the cavity mode, as shown in Fig.~\ref{angle}. This can also be understood in terms of the heavy mass of the excitons when compared with the photons' effective mass. The formation of the optical mode relies on the excitation of electron-hole pairs; this implies that the photon should be resonant with the excitonic line for all the momenta of the cavity dispersion. Therefore, the cavity inherits its dispersion from the excitonic momentum distribution, which within the light cone, is flat.

%%%%%%%%%%%%%%%%%%%%%%%%%%%%%%%%%%%%%%%%%%%%%%%%%%%%%%%%%%%%%%%%%%%%%%%%%

\begin{figure}[ht]
    \centering
    \includegraphics[width=\columnwidth]{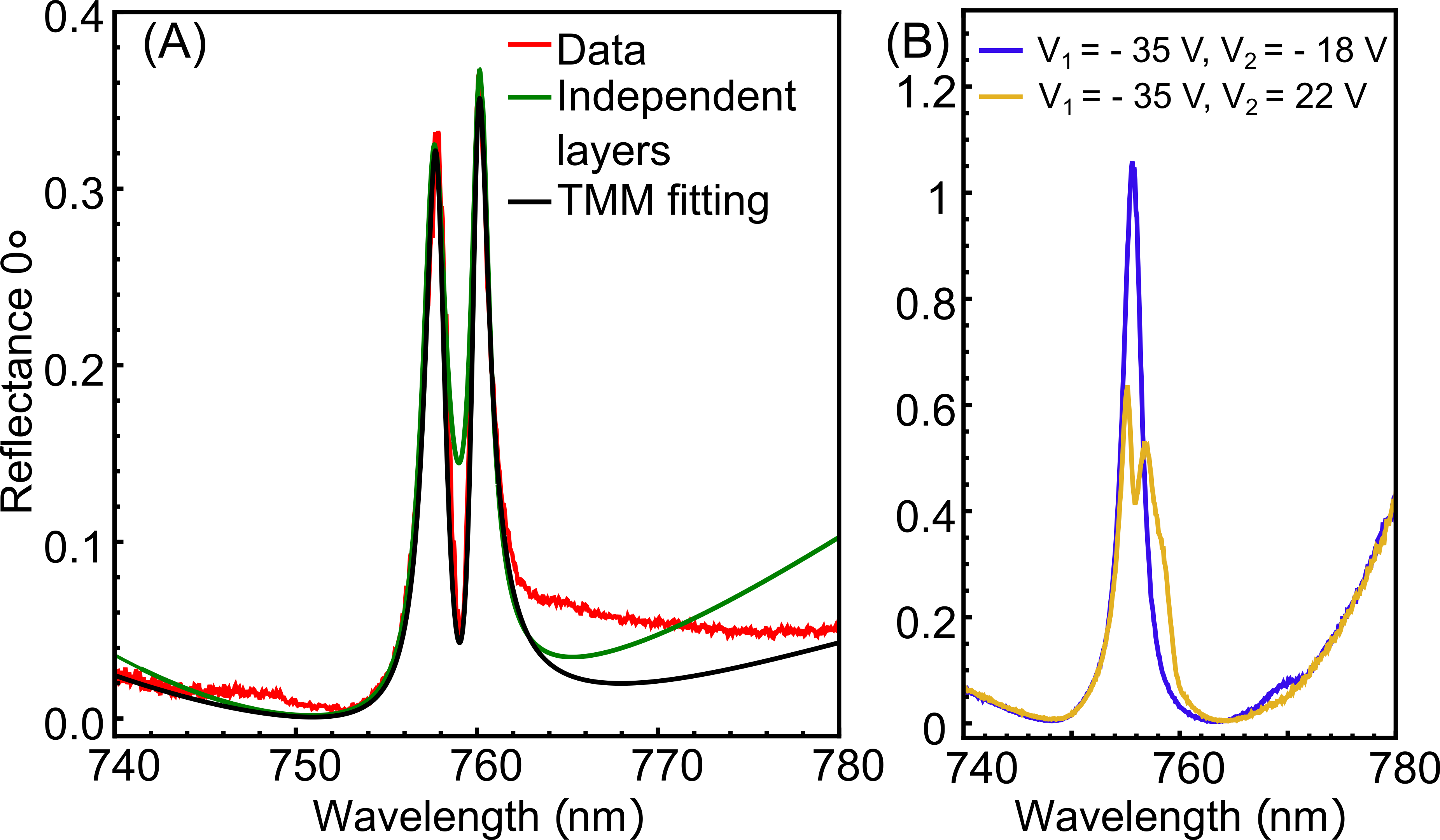}
    \caption{\textbf{Identification of cavity mode vs non-interacting oscillators.} (A) Experimental data at the spot where the magnetic field dependence was measured (red line) fitted with a model where the two TMD layers are decoupled (green line). The impossibility of fitting the experimental reflectance with this model in comparison with the TMM model that considers the full structure (black line) confirms that the observed dip in the reflectivity corresponds to an optically confined mode instead of two independent excitonic reflectors. (B) The electrical control permits quenching the excitonic response of one of the monolayers (blue line). When both monolayers have excitonic resonances, the spectrum does not correspond to the sum of two Lorentz oscillators, but the cavity mode emerges instead (yellow line).}
    \label{lorentz-vs-tmm}
\end{figure}

\subsection{\label{si:8} 8. Identification of cavity mode vs non-interacting oscillators}

\begin{figure*}
    \centering
    \includegraphics[width=1.5\columnwidth]{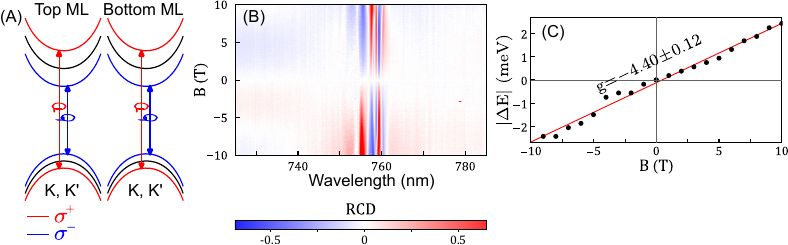}
    \caption{\textbf{Additional data: magnetic field-induced chirality.} (A) Schematic showing the Zeeman splitting in top and bottom monolayers. This gives rise to the two chiral modes of the cavity with $\sigma^+$ and $\sigma^-$ polarizations. (B) RCD of the heterostructure cavity for increasing $B$ at a different spot. (C) Energy difference $\Delta E$ between the $\sigma^+$ and $\sigma^-$ cavity modes as a function of $B$. From a linear regression, we extract a  magnetic factor $g=-4.40\pm0.12$}
    \label{magnet_SM}
\end{figure*}

In some spots of the sample, the identification of the cavity mode is not straightforward, because it might be confused with the reflection from two decoupled Lorentz oscillators. To confirm that the observed spectrum corresponds to a TMD cavity, we fit the spectrum using a model with two independent excitonic resonances, and we compare the result with the obtained reflectance from the full structure. As observed in Fig.~\ref{lorentz-vs-tmm}, the independent oscillators model cannot properly fit the experimental data, in contrast to the TMM fitting that properly accounts for the measured spectrum. This spectrum corresponds to the sample spot where the magnetic field dependence was collected (Fig.~3 of the main text). In panel B of the same figure, we show additional evidence of the formation of genuine optical modes in the structure. Upon the quenching of the exciton resonance in one of the monolayers via an electrical control, the spectrum corresponds to the one of a MoSe$_2$ mirror. When both excitons are available, the spectrum does not correspond to the trivial addition of them, but the interference and hence the cavity mode is observed.

%%%%%%%%%%%%%%%%%%%%%%%%%%%%%%%%%%%%%%%%%%%%%%%%%%%%%%%%%%%%%%%%%%%%

\subsection{\label{si:9} 9. Additional data: magnetic field-induced chirality}

Due to high spatial inhomogeneity in exfoliated TMDs, the excitonic resonance frequency, $\Gamma_{\text{r}}$ and $\Gamma_{\text{nr}}$ can vary notably across different spots on the sample. However, we observe the formation of a cavity, albeit with slightly different resonant energies and $\Gamma_{\text{r}}/\Gamma_{\text{nr}}$ ratios, across a substantial portion of the sample.

We performed additional magnetically induced reflective circular dichroism (CD) measurements over a larger range of magnetic fields in another spot on the sample for comparison of different cavity spots. The magnetic field varies from -10 T to 10 T. Fig. \ref{magnet_SM}B shows the reflective CD measurement results. For the highest applied magnetic field, we obtain a contrast of $>1$ in the CD, which is larger than our measurement in the original sample spot.
From the magnetically induced energy splitting of the chiral modes $\Delta E$ of the cavity we extract a g-factor of $g\!=\!-4.40\pm0.12$ (panel C), using the relationship $\Delta E = g\mu_B B$, where $\mu_B$ is the Bohr magneton and $B$ is the magnetic field strength. The obtained value is in good agreement with the reported value in the main text (and in the literature). 

While the requirement for a high magnetic field would not be a restriction for any of the envisaged perspectives, it is important to relax this constraint to enlarge the list of possible applications of this architecture. One option is to harness the demonstrated valley optical Stark effect to induce a chiral behavior without the need for a magnetic field \cite{lamountain2021valley,hao2022optically}. Another possibility to reduce the required magnetic field is to interface the device with magnetic materials. The magnetic exchange field \cite{zhao2017enhanced} and the magnetic proximity effect \cite{norden2019giant} are reportedly effective in substantially enhancing the magnetic response of the excitons in TMD materials.

%%%%%%%%%%%%%%%% SUPPLEMENTARY TABLES %%%%%%%%%%%%%%%

\begin{table}[ht]
	\centering
	% Captions go above tables
	\caption{\textbf{Q-factors of planar cavities reported in the literature.}}
	\label{tab:Q} % give each table a logical label name
\begin{tabular}{ccc} % four columns, alignment for each
		\\
		\hline
		Type of mirrors & Q-factor & Reference \\
		\hline
		DBRs & ~500 & \cite{anton2021bosonic} \\
		DBRs & 400-700 & \cite{peng2024topological} \\
        DBRs & ~3700 (median) & \cite{paik2023high} \\
        DBRs & 100-500 & \cite{pandya2022tuning} \\
        DBRs & 600 & \cite{sannikov2019room} \\
        DBRs & 440-620 & \cite{connolly2003strong} \\
        DBRs & 300 & \cite{virgili2011ultrafast} \\
        DBR + metallic & 1000-3000 & \cite{mcghee2021polariton} \\
        Metallic + grating mirrors & 100-1000 & \cite{byrnes2016high} \\
        Metallic & 35 & \cite{coles2017strong} \\
		\hline
	\end{tabular}
\end{table}

%%%%%%%%%%%%%%%%%%%%%%%%%%%%%%%%%%%%%%%%%%%%%%%%%%%%%%%%%%%%%%%%%%%%%%%%%%%%%%%%%%%%%%%%%%%%%%%%%%%%%%%%%%%%%%%%%%%%%%%%%%%%%%

\end{document}

%% file: authors.tex
\author{Daniel G. Suárez-Forero}
\altaffiliation{These authors contributed equally to this work}
\affiliation{Joint Quantum Institute (JQI), University of Maryland, College Park, MD 20742, USA}

\author{Ruihao Ni}
\altaffiliation{These authors contributed equally to this work}
\affiliation{Department of Materials Science and Engineering, University of Maryland, College Park, MD 20742, USA}

\author{Supratik Sarkar}
\altaffiliation{These authors contributed equally to this work}
\affiliation{Joint Quantum Institute (JQI), University of Maryland, College Park, MD 20742, USA}

\author{Mahmoud Jalali Mehrabad}
\altaffiliation{These authors contributed equally to this work}
\affiliation{Joint Quantum Institute (JQI), University of Maryland, College Park, MD 20742, USA}

\author{Erik Mechtel}
\affiliation{Joint Quantum Institute (JQI), University of Maryland, College Park, MD 20742, USA}

\author{Valery Simonyan}
\affiliation{Joint Quantum Institute (JQI), University of Maryland, College Park, MD 20742, USA}

\author{Andrey Grankin}
\affiliation{Joint Quantum Institute (JQI), University of Maryland, College Park, MD 20742, USA}

\author{Kenji Watanabe}
\affiliation{Research Center for Electronic and Optical Materials, National Institute for Materials Science, 1-1 Namiki, Tsukuba 305-0044, Japan
}

\author{Takashi Taniguchi}
\affiliation{Research Center for Materials Nanoarchitectonics, National Institute for Materials Science,  1-1 Namiki, Tsukuba 305-0044, Japan}

\author{Suji Park}
\affiliation{Center for Functional Nanomaterials, Brookhaven National Laboratory, Upton, NY 11973, USA}

\author{Houk Jang}
\affiliation{Center for Functional Nanomaterials, Brookhaven National Laboratory, Upton, NY 11973, USA}

\author{Mohammad Hafezi$^{\dagger}$}
\affiliation{Joint Quantum Institute (JQI), University of Maryland, College Park, MD 20742, USA}

\author{You Zhou$^{\ddag}$}
\affiliation{Department of Materials Science and Engineering, University of Maryland, College Park, MD 20742, USA}
\affiliation{Maryland Quantum Materials Center, College Park, Maryland 20742, USA}